\documentclass[aps,prl,showpacs,twocolumn,superscriptaddress]{revtex4}
\usepackage{bm}
\usepackage[dvipdfmx]{graphicx}
\usepackage[dvipdfmx]{color}
\usepackage{amsmath, amssymb}
\usepackage{braket}
\usepackage{color}
\usepackage{comment}

\begin{document}
\title {Theory of magnetic vortex crystals induced by electric dipole interactions}
\author{Tatsuki Muto}
\affiliation{Department of Applied Physics, Waseda University, Okubo, Shinjuku-ku, Tokyo 169-8555, Japan}
\author{Masahito Mochizuki}
\affiliation{Department of Applied Physics, Waseda University, Okubo, Shinjuku-ku, Tokyo 169-8555, Japan}
\begin{abstract}
Recent Lorentz transmission electron microscopy experiments have revealed a variety of noncollinear spin textures including topological magnetisms such as skyrmions, biskyrmions and multiple skyrmions in thin-plate magnetic samples of e.g., perovskite manganites and hexaferrites whose crystal structures originally have spatial inversion symmety. Motivated by these experiments, we investigate a spin-lattice coupled model involving both magnetic and electric dipole interactions by the Monte Carlo technique based on the stochastic cut-off method and propose a physical mechanism for realizing topological spin textures in which the electric dipole interaction plays a key role. This mechanism is ubiquitous that stabilizes noncollinear and topological spin textures in thin-plate samples of magnetic materials.
\end{abstract}
\maketitle

\section{Introduction}
Topological spin textures are recently attracted enormous interest as a treasure trove of novel physical phenomena and material functions~\cite{Seki16,Seidel16,Nagaosa13,Gobel21,Fert13,Finocchio16,Fert17,Everschor19,Tokura21}. Skyrmion is a particle composed of vector fields pointing in all directions covering a sphere and was originally proposed by Tony Skyrme in 1960s as a theoretical model for baryons in particle physics~\cite{Skyrme61,Skyrme62}. After more than a decade, Bogdanov and coworkers theoretically predicted that skyrmions could be manifested as spin textures in magnets~\cite{Bogdanov89,Bogdanov94,Bogdanov99,Rossler06}. Recently, small-angle neutron scattering experiments and Lorentz transmission electron microscopy experiments have revealed that they are indeed realized in magnetic materials with chiral crystal structures such as B20 compounds (e.g., MnSi, Fe$_{1-x}$Co$_x$Si, FeGe)~\cite{Muhlbauer09,YuXZ10,Munzer10,YuXZ11,Tonomura12} and Cu$_2$OSeO$_3$~\cite{Seki12a,Seki12b,Seki12c,Adams12}. The skyrmions in these magnets are circular-shaped vortex-like spin textures, which are called Bloch-type. The spatial spin configuration of the skyrmion can be regarded as a stereo-projection of the originally proposed three-dimensional skyrmion onto a two-dimensional plane. The skyrmions in magnets appear not only as isolated defects in the ferromagnetic state but also in the crystallized form called skyrmion crystals~\cite{YuXZ10,YuXZ11}.

The crystal structures of the B20 compounds and Cu$_2$OSeO$_3$ have a chiral cubic symmetry with broken spatial inversion symmetry, which activates the DM interaction~\cite{Dzyaloshinsky58,Moriya60a,Moriya60b}. This interaction favors a helical spin configuration and thus competes with the ferromagnetic exchange interaction that favors a parallel spin configuration. When an external magnetic field of a certain intensity is applied to this system,  skyrmion crystals or isolated skyrmions appear. The simplest Hamiltonian describing such a situation in a chiral magnet hosting skyrmions, therefore, consists of three interactions: the ferromagnetic exchange interaction, the DM interaction, and the Zeeman interaction associated with the external magnetic field~\cite{Bak80,YiSD09}. The type of skyrmion is governed by the structure of DM vectors determined by the crystal structure. In addition to the Bloch-type skyrmions in chiral magnets, the Neel-type skyrmions with fountain-shaped spin configurations have been found in polar magnets, e.g., GaV$_4$S$_8$~\cite{Kezsmarki15,Ruff15}, GaV$_4$Se$_8$~\cite{Fujima17,Akazawa22} and VOSe$_2$O$_5$~\cite{Kurumaji17,Kurumaji21}, and the antivortex-type skyrmions with a negative spin vorticity have been found in a Huesler compound Mn$_{1.4}$Pt$_{0.9}$Pd$_{0.1}$Sn~\cite{Nayak17} and Fe$_{1.9}$Ni$_{0.9}$Pd$_{0.2}$P~\cite{Karube21}.

Because of this historical reason, the magnetic skyrmions have long been considered to appear only in magnets  with broken spatial inversion symmetry such as chiral magnets and polar magnets. However, in recent years, skyrmions and skyrmion crystals have been discovered in magnets having spatial inversion symmetry, in which the DM interaction is expected to be absent. For example, a variety of skyrmion crystals and other topological magnetic structures have been discovered in bulk magnetic materials (e.g., SrFeO$_3$~\cite{Ishiwata20}, Gd$_2$PdSi$_3$~\cite{Kurumaji19,Hirschberger20a,Hirschberger20b,Nomoto20}, Gd$_3$Ru$_4$Al$_{12}$~\cite{Hirschberger19,Hirschberger21,Khanh20,Yasui20}) and thin-plate samples of M-type hexaferrites~\cite{YuXZ12} and perovskite Mn oxides~\cite{YuXZ14,Nagao13}. Theoretically, several mechanisms that realize skyrmions without DM interaction have been proposed, originating from geometrical spin frustrations~\cite{Okubo12,Kamiya14,Leonov15,LinSZ16,Hayami16a,Hayami16b,LinSZ18,Lohani19,WangZ21,Kharkov17}, long-range spin interactions mediated by conduction electrons~\cite{Hayami21R,Akagi12,Hayami14b,Hayami14a,Ozawa16,Ozawa17a,Hayami17,Gobel17,Wang20,Hayami20a,Hayami21a,Hayami21b,Nakazawa19,Hayami20b}, and magnetic dipole interactions~\cite{Correspondent72,LinYS73,Malozemoff79,Giess80,Garel82,Suzuki83,Hubert98}. The former two mechanisms have been discussed, in particular, as the origin of the skyrmion-crystal formation in bulk magnetic materials with spatial inversely symmetry.

On the other hand, several nontrivial spin textures including skyrmions, biskyrmions (i.e., two linked skyrmions), and modulated helimagnetism have been observed in thin-plate samples of magnets\cite{YuXZ12,YuXZ14,Nagao13}. They have been naively considered to originate from magnetic dipole interactions. However, the nanometre-sized skyrmion-like spin textures observed in this system are clearly different in size and spatial spin configuration from the micrometre-sized magnetic bubbles naively expected in systems with magnetic dipole interactions, suggesting the existence of different mechanism. 

Even in magnets with spatial inversion symmetry, local symmetry breaking due to the displacement of ion is expected to result in emergence of the local electrical polarization and the local DM vectors. Furthermore, the local electric polarizations should couple with noncollinear spins via the local DM interaction. In thin-plate  samples, the spatial structure of the DM vectors derived from the electric polarization is governed by the  dipole interaction, while the noncollinear spin configuration  is governed by the magnetic dipole interaction, resulting in a formation of noncollinear configurations of the spins and the polarizations, even though the original material has a spatial inversion symmetry. Such noncollinear spin configurations can often be topological spin textures such as skyrmions, biskyrmions and multiple skyrmions.

In this paper, we numerically analyze a spin-lattice model describing the coupling between the spins and the local electric polarizations in thin-plate magnets by utilizing the Monte Carlo technique based on the stochastic cut-off method~\cite{Sasaki08}. Through this analysis, we propose a physical mechanism for realizing noncollinear and even topological magnetisms in which the electric dipole interaction plays a key role. This mechanism is expected to be universal in thin-plate magnetic systems, and various nontrivial spin textures can be expected through varying the strength of spin-lattice coupling and/or through tuning competition between the electric and magnetic dipole interactions, which depend, e.g., on materials and sample thickness. Therefore, our proposal and findings pave the way to the topology engineering using thin-plate magnets.

\section{Model and Method}
\begin{figure}[tb]
\includegraphics[scale=1.0]{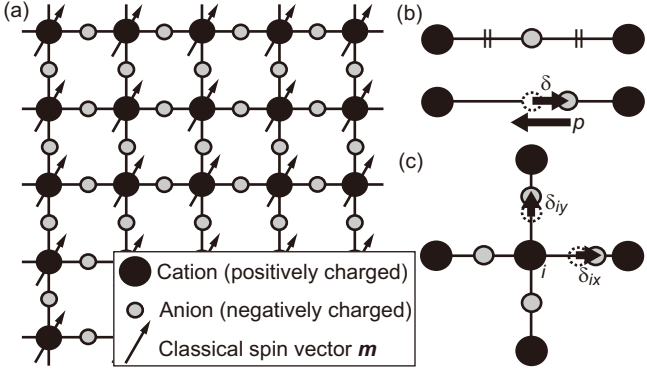}
\caption{(a) Two-dimensional plane with cation sites $M$ and anion sites $L$ considered in the present work. The cation sites with a classical spin $\bm m_i$ constitute a square lattice, while the ligand anions are located in-bewteen the adjacent  cations. (b) The anions $L$ are originally located at the center of the $M$-$M$ bonds, while their displacements generate the local electric polarization. (c) Assignment and definition of the displacement vector $\bm \delta_i=(\delta_{ix}, \delta_{iy})$ of the $ith$ cation $M_i$.}
\label{Fig01}
\end{figure}
To describe the spin-lattice coupled system in a thin-plate magnet, we employ a classical spin-lattice model on a square lattice (see Fig.~\ref{Fig01}). The Hamiltonian is composed of three contributions, i.e., the spin term, the lattice term, and the spin-lattice coupling term as,
\begin{equation}
\mathcal{H}=\mathcal{H}_{\rm spin}+\mathcal{H}_{\rm lattice} +\mathcal{H}_{\rm sl}.
\label{eq:Hamlt}
\end{equation}
The spin term $\mathcal{H}_{\rm spin}$ is given by,
\begin{align}
\mathcal{H}_{\rm spin}&=-J\sum_{<i,j>}\bm m_i \cdot \bm m_j 
\nonumber \\
&+D_{\rm m}\sum_{i<j} \left[ \frac{\bm m_i \cdot \bm m_j}{r_{ij}^3}
-\frac{3(\bm m_i \cdot \bm r_{ij})(\bm m_j \cdot \bm r_{ij})}{r_{ij}^5} \right] 
\nonumber \\
&-A\sum_{i} m_{iz}^2 - H_z\sum_i m_{iz},
\end{align}
where $\bm m_i$ is a classical spin vector at the $i$th site whose norm is unity, i.e., $|\bm m_i|=1$. The first term describes the ferromagnetic exchange interactions, while the second term describes the magnetic dipole-dipole interaction. The summation $\sum_{<i,j>}$ is taken over the adjacent site pairs. The third term represents the pependicular magnetic anisotropy with $A>0$. 

On the other hand, the lattice term $\mathcal{H}_{\rm lattice}$ is given by,
\begin{align}
\mathcal{H}_{\rm lattice}&=D_{\rm e}\sum_{i<j} \left[ \frac{\bm \delta_i \cdot \bm \delta_j}{r_{ij}^3}
-\frac{3(\bm \delta_i \cdot \bm r_{ij})(\bm \delta_j \cdot \bm r_{ij})}{r_{ij}^5} \right] 
\nonumber \\
&+K\sum_i\left(\delta_{ix}^2+\delta_{iy}^2\right).
\end{align}
Here $\delta_{i\mu}$ denotes the displacement of ligand anion on the bond between the $i$th and ($i$+$\hat{\mu}$)th  cations  along the $\mu$ direction ($\mu$=$x$, $y$), which is normalized by the original bond length [Fig.~\ref{Fig01}(a)]. The displacements give rise to local electric polarizations, and the polarization vector $\bm p_i$ that belongs to the $i$th site is defined as $\bm p_i  \propto \bm \delta_i=(\delta_{ix}, \delta_{iy})$. The first term describes the electric dipole-dipole interactrions among the local polarizations, while the second term describes the elastic energy due to the displacement of ligand ions from their original positions.

The displacements of ligand ions break the spatial inversion symmetry at the bonds, which locally activate the DM interactions beween the adjacent spins. We assume that the strength of the local DM interaction proportionally scales with the displacement or the local polarization. Accordingly, the spin-lattice coupling in this thin-plate system is described by,
\begin{equation}
\mathcal{H}_{\rm sl}=\alpha \sum_i \left[ \delta_{ix}(\bm m_i \times \bm m_{i+\hat{x}})_x
+\delta_{iy}(\bm m_i \times \bm m_{i+\hat{y}})_y \right],
\end{equation}
where $\alpha$ is the coupling constant.

We analyse this model by using the Monte-Carlo technique for square-lattice systems of $N=48\times48$ sites and $N=72\times72$ sites with periodic boundary conditions. Both the spin vectors $\bm m_i$ and the displacements $\bm \delta_i$ are updated in the Monte-Carlo procedure. Because the magnetic and electric dipole-dipole interactions are long-ranged, we need to take not only the adjacent site pais but also all the further site pairs into account, which inevitably results in huge computational time of $O(N^2)$, if we use a simple Monte-Carlo technique. In the present work, we employ the stochastic cut-off method. This method can reduce the calculation cost by stochastically selecting bonds on which the interactions are considered for each Mote-Carlo step, while satisfying the detailed balance condition.  

For demonstrations, we set the model parameters as $J=1$, $D_{\rm m}=0.2$, $A=1.45$, $D_{\rm e}=150$ and $K=500$. We search low-energy spin-lattice configurations by gradually decreasing temperature from a higher temperature of $k_{\rm B}T/J=1$ to a lower temperature of $k_{\rm B}T/J=0.01$ starting from a random spin configuration. We examine many random configurations as initial states and compare energies of the final states to determine the ground-state and metastable spin-lattice configurations. In the Monte-Carlo procedure, we generate new displacements $\delta_{ix}$ and $\delta_{iy}$ within  a certain window of $-\delta_{\rm w\mu} \leq \delta_{i\mu} \leq \delta_{\rm w\mu}$ ($\mu$=$x$, $y$) to update them by generating random numbers. We set the value of $\delta_{\rm w\mu}$ for the $k$th Monte-Carlo step ($k \ge 1$) as $\delta_{\rm w\mu}=|\bar{\delta}_\mu|+3\sigma_\mu$. Here $\bar{\delta}_\mu \equiv \sum_i \delta_{i\mu}/N$ is the site-averaged displacement, and $\sigma_\mu$ is the standard deviation of  $\delta_{i\mu}$ for the $(k-1)$th step. We set  $\delta_{\rm w\mu}=0.15$ for the first step ($k=0$).

\section{Results}
\begin{figure}[tb]
\includegraphics[scale=1.0]{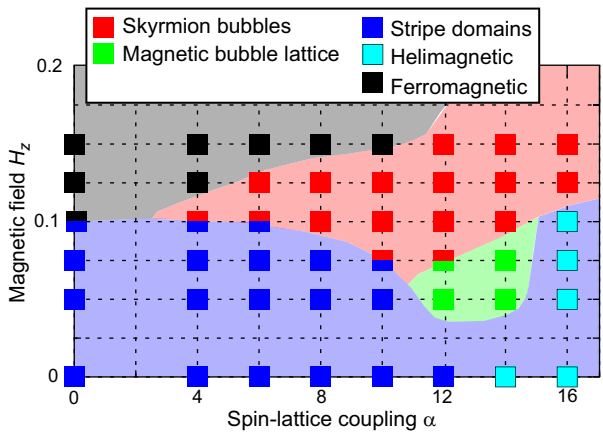}
\caption{Ground-state phase diagram of the spin-lattice model in Eq.~(\ref{eq:Hamlt}) in plane of the spin-lattice coupling $\alpha$ and the external magnetic field $H_z$.}
\label{Fig02}
\end{figure}
\begin{figure}[tb]
\includegraphics[scale=1.0]{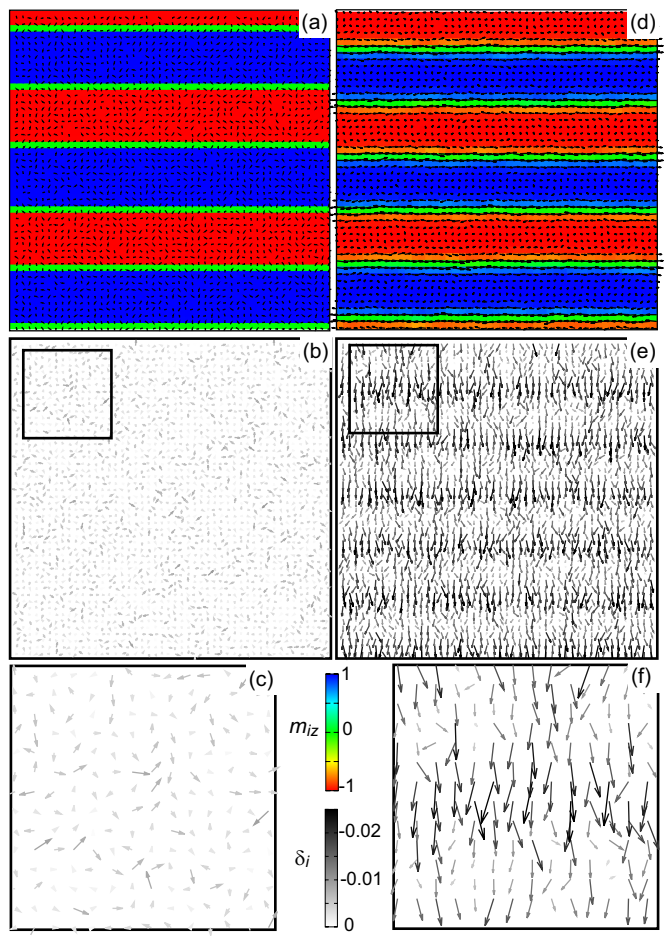}
\caption{(a)-(c) Spatial configurations of (a) spins and (b) polarizations for the stripe-domain state in the absense of the spin-lattice coupling ($\alpha$=0) when $H_z=0$. The area indicated by the solid square in (b) is magnified in (c). (d)-(f) Those of the helimagnetic state in the presence of the spin-lattice coupling ($\alpha$=14) when $H_z=0$. The uniform component of polarization appears parallel to the helimagnetic propagation vector or perpendicular to the stripe.}
\label{Fig03}
\end{figure}
\begin{figure}[tb]
\includegraphics[scale=1.0]{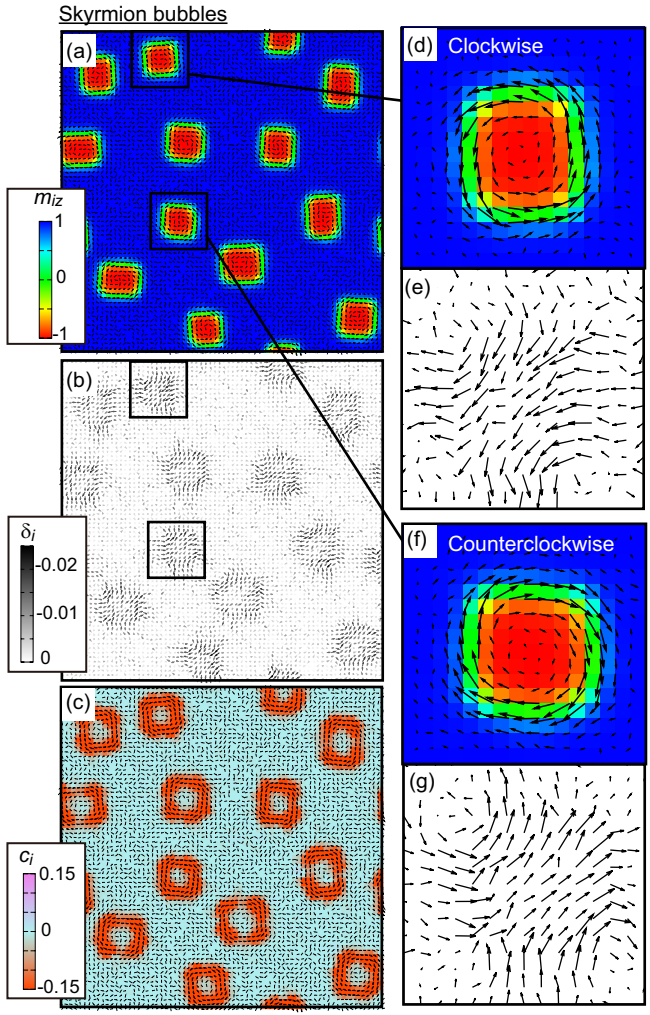}
\caption{(a)-(c) Spatial configurations of the (a) spins, (b) polarizations, and (c) local scalar spin chiralities of the skyrmion bubbles in a system of 72 $\times$ 72 sites with $\alpha=14$ and $H_z=0.125$. (d), (e) Enlarged views of the (d) spin and (e) polarization configurations of a counterclockwise skyrmion bubble. (f), (g) Those of a clockwise skyrmion bubble. }
\label{Fig04}
\end{figure}
\begin{figure}[tb]
\includegraphics[scale=1.0]{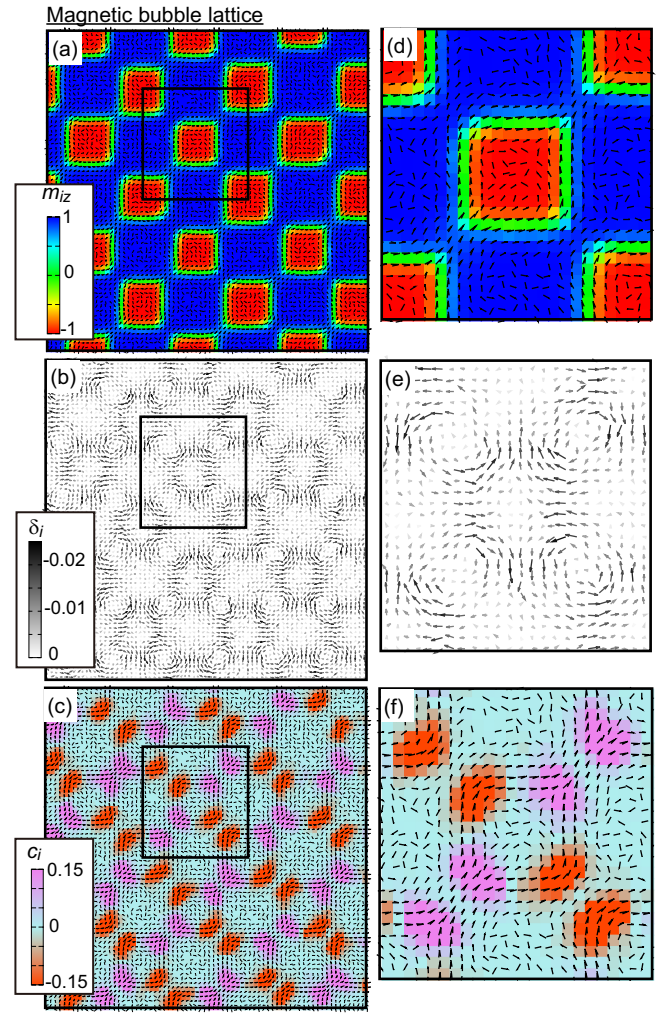}
\caption{(a)-(c) Spatial configurations of the (a) spins, (b) polarizations, and (c) local scalar spin chiralities of the magnetic bubble lattice in a system of 72 $\times$ 72 sites with $\alpha=14$ and $H_z=0.075$. (d)-(f) Enlarged views of the areas indicated by the solid squares in (a)-(c). }
\label{Fig05}
\end{figure}
\begin{figure}[tb]
\includegraphics[scale=1.0]{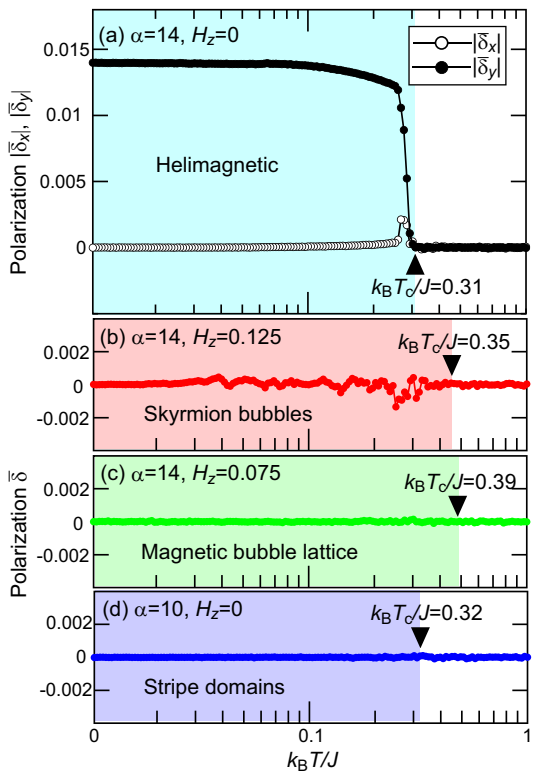}
\caption{(a),~(b) Temperature profiles of the (a) average and (b) variance of anion displacements $\delta_i$ over sites ($i=1, 2, \cdots ,N$ ) for the helimagnetic phase with $\alpha=14$ and $H_z=0$. When the system enters the low-temperature helimagnetic phase with decreasing temperature, both the average and variance start increasing. (c)-(e) Temperature profiles of the site-averages of $\delta_i$ which exhibit no anomaly upon  the phase transitions to the (c) skyrmion-bubble state, (d) magnetic bubble-lattice state and (e) stripe-domain state, resepectively.}
\label{Fig06}
\end{figure}
\begin{figure*}[tb]
\includegraphics[scale=1.0]{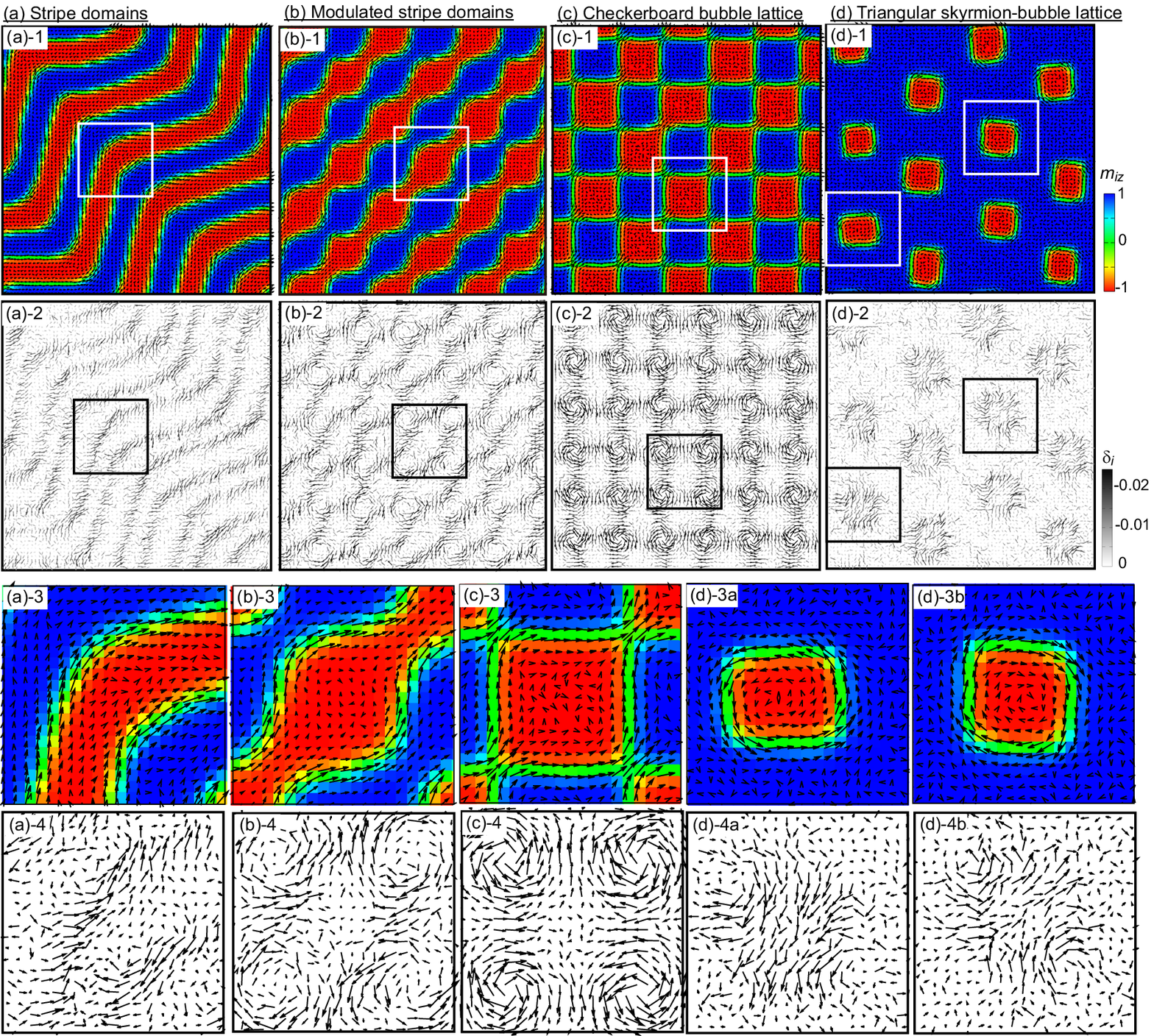}
\caption{Spin and polarization configurations of various metastable states: (a) the stripe domains for $\alpha=10$ and $H_z=0$, (b) the moulated stripe domains for $\alpha=10$ and $H_z=0$, (c) the checkerboard bubble lattice for $\alpha=12$ and $H_z=0$, and (d) the triangular skyrmion bubble lattice for $\alpha=10$, $H_z=0.1$ and $D_{\rm e}=160$. The panels in the first (second) law show spin (polarization) configurations in a system of 72 $\times$ 72 sites. The panels of the third (fourth) law show enlarged views of the areas indicated by the solid squares in the panels of the first (second) law.}
\label{Fig07}
\end{figure*}
We first discuss the ground-state phase diagram in plane of the spin-lattice coupling $\alpha$ and the external magnetic field $H_z$ (Fig.~\ref{Fig02}). When the spin-lattice coupling is absent or weak as $\alpha<4$, the stripe-domain state [Fig.~\ref{Fig03}(a)] appears at $H_z=0$. This stripe-domain state survive even under application of the magnetic field up to $H_z=0.1$. With further increasing $H_z$, the system enters the field-polarized ferromagnetic state. The stripe domains are stabilized by the magnetic dipole interactions, which are not accompanied by the anion displacements or electric polarizations as shown in Figs.~\ref{Fig03}(b) and (c), where the electric dipole interaction does not play any substantial roles. Note that neither topological magnetism nor magnetic vortices appear at least for the present parameter set. It should also be mentioned that the spin orientation is suddenly reversed from upwards to downwards or vice versa at the atomically thin domain walls in the present system with a strong perpendicular anisotropy. 

As $\alpha$ increases, this stripe-domain state changes into the helimagnetic state at $\alpha \sim 13$. This helimagnetic state is characterized by gradual spatial variation of spin configuration instead of the sudden reversal at the atomically thin domain walls in the stripe-domain state [Fig.~\ref{Fig03}(d)]. This gradual spin rotation is caused by the Dzyaloshinskii-Moriya interactions due to the local anion displacements induced via the spin-lattice coupling as shown in Figs.~\ref{Fig03}(e) and (f). In the helimagnetic state, the anion displacements occur in the same direction resulting in the emergence of ferroelectric polarization. This helimagnetic state is stabilized dominantly by the magnetic dipole interaction because the uniform electric polarization is energetically unfavorable for the electric dipole interaction.

The most interesting finding of the present work is that the skyrmion bubble state [Fig.~\ref{Fig04}(a)] appears when the spin-lattice coupling is enough strong under application of the magnetic field as an intermediate state between the stripe-domain/helimagnetic state and the ferromagnetic state. These skyrmion bubbles are accompanied by the anion displacements as shown in Fig.~\ref{Fig04}(b). There are two kinds of skyrmion bubbles with respect to the spin helicity, that is, clockwise and counterclockwise vortex configurations [Figs.~\ref{Fig04}(d) and (f)]. They are both accompanied by the anion displacements, and their directions are opposite to each other. Because the system originally has spatial inversion symmetry, these two types of skyrmion bubbles are degenerate. 

We calculate the local scalar spin chirality $C_i$, which is given by,
\begin{equation}
C_i=\bm m_i \cdot (\bm m_{i+\hat{\bm x}}\times \bm m_{i+\hat{\bm y}})
+\bm m_i \cdot (\bm m_{i-\hat{\bm x}}\times \bm m_{i-\hat{\bm y}}).
\end{equation}
Its spatial distribution in Fig.~\ref{Fig04}(c) indicates that both the clockwise and counterclockwise skyrmion bubbles have finite negative scalar spin chirality at their peripheral areas with noncollinear spin alignment. On the contrary, the area around each core has zero scalar spin chirality, indicating that the uniform or collinear spin configuration is realized there. This is why we call these spin textures skyrmion bubbles instead of skyrmions. We also find that the summation of $C_i$ over sites constituting one skyrmion bubble takes $-4\pi$, which corresponds to a quantized topological charge of $Q=-1$. Importantly, these topological spin textures appear only when the spin-lattice coupling of sufficient strength (i.e., $H_z>4$) is present. 

We observe another nontrivial spin state [Figs.~\ref{Fig05}(a) and (b)] within a small area of finite $\alpha$ and $H_z$ in the phase diagram, which we refer to as the magnetic bubble lattice. In this state, square-shaped bubble domains are aligned to form a square lattice. The spins constituting a peripheral area of each bubble domain are aligned in a noncollinear manner, but they do not form a vortex. These spin textures are accompanied by whirling configurations of local anion displacements or vortex formations of electric polarizations. Interestingly, we observe a staggered alignment of vortieces with opposite senses, that is, the clockwise and counterclockwise vortices of polarizations are aligned alternatingly to form a square lattice. These polarization vortices and their staggered arrangement are favored by the electric dipole interactions.

It should be mentioned that these magnetic bubbles are topologically trivial spin textures. In Figs.~\ref{Fig05}(c) and (f), we plot the calculated spatial distribution of local scalar spin chirality $C_i$. We find that in each magnetic bubble, there are portions with opposite signs of chirality, and they cancel out each other, resulting in vanishing of the net scalar spin chirality. This spatially structured local spin chirality indicates the existence of spatially modulated emergent magnetic field. This might causes some nontrivial emergent physical phenomena, which should be clarified in future studies. 

In Fig.~\ref{Fig06}, we plot temperature profiles of the net polarization $\bar{\delta}(\equiv \sqrt{\bar{\delta}_x^2+\bar{\delta}_y^2})$ for several parameter sets with which certain spin states appear at low temperatures. The components $\bar{\delta}_x$ and $\bar{\delta}_y$ are calculated by summing up the local contributions as,
\begin{equation}
\bar{\delta}_\mu=\frac{1}{N} \sum_i \delta_{i\mu} \quad (\mu=x,\;y),
\end{equation}
where $N$ is the total number of sites. We find that the absolute value $\bar{\delta}$ abruptly increases at the transition to the helimagnetic state in Fig.~\ref{Fig06}(a). On the contrary, $\bar{\delta}$ remains to be zero upon the transitions to the skyrmion bubbles, the magnetic bubble lattice, and the stripe domains. Importantly, the skyrmion bubbles and the magnetic bubble lattice do not have uniform component of the anion displacements or ferroelectric polarization even though these two states apparently have nonzero local polarizations as seen in Fig.~\ref{Fig04} and Fig.~\ref{Fig05}. This indicates that in these states, the electric polarizations are distributed so as to close the eletric flux lines, favored by the electric dipole interactions. 

We finally show several metastable spin states obtained in the present calculations in Fig.~\ref{Fig07}. In the simulation, we gradually decrease temperature from higher to lower starting with a random spin and polarization configurations. Among many trials with different initial random configurations, we often encounter metastable spin-polarization configurations which are trapped by a local energy minimum. These metastable states may be realized/observed in real experiments depending on the temperature and magnetic field sweep history. Figures~\ref{Fig07}(a)-(d) show spatial configurations of spins and anion displacements for the winding stripe-domain state, the modulated stripe-domain state, the checkerboard bubble lattice, and the triangular skyrmion-bubble lattice. We find that these noncollinear spin textures are accompanied by a specific spatial structure of the polarizations involving polarization-vortex crystals.

\section{Summary}
In summary, we have theoretically demonstrated a general mechanism for realizing noncollinear and topological textures of spins and polarizations in thin-plate magnets which originally have a centrosymmetric crystal structure by analyzing the spin-lattice model on a square lattice. The model involves both magnetic and electric dipole interactions, and we employ the Monte Carlo technique based on the stochastic cut-off method, which is known as a poweful numerical method to treat the systems with long-range interactions. The local displacements of ligand ions generate local electric plarizations and thus break the local inversion symmetry, which activate the DM interactions locally. The induced local DM interactions mediate the coupling between the spins and the polarizations.  We reveal that the mutual coupling beween the spin and polarization configuration stailized by the magnetic and electric dipole interactions give rise to their rich textures. This mechanism is a general one, which can be expected in a wide variety of magnetic materials. We anticipate that the proposed mechanism will be a key to engineering the magnetic topology in thin-plate magnets for future spintronics application.

\section{Acknowledgment}
This work is supported by 
Japan Society for the Promotion of Science KAKENHI (Grants No. 20H00337 and No. 22H05114), 
CREST, the Japan Science and Technology Agency (Grant No. JPMJCR20T1), and 
a Waseda University Grant for Special Research Projects (Projects No. 2023C-140).

\end{document}